%Paper: solv-int/9304003
%From: Robert I McLachlan <rxm@vortex.Colorado.EDU>
%Date: Wed, 28 Apr 93 13:58:33 -0600

\hsize=5.5in
\hoffset=0.5in
\def\p{{\bf p}}
\def\x{{\bf x}}
\def\y{{\bf y}}
\def\f{{\bf f}}
\def\u{{\bf u}}
\def\g{{\bf g}}
\def\h{{\bf h}}
\def\eps{\varepsilon}
\font\bigb=cmbx12
\font\big=cmr12
\hbox{\ }\vskip 0.7in
\centerline{\bigb Integrable four-dimensional symplectic maps of standard type}
\bigskip
\centerline{\big Robert I. McLachlan}
\bigskip \bigskip
\centerline{\it Program in Applied Mathematics, University of Colorado at
Boulder, Boulder, CO 80302}
\bigskip\bigskip
\noindent (To appear in {\sl Phys. Lett. A})
\vskip 1in\centerline{\bf Abstract}\bigskip
{\narrower\noindent
We search for rational, four-dimensional maps of standard type
($\x_{n+1}-2\x_n+x_{n-1}=\eps\f(\x,\eps)$) possessing one or two polynomial
integrals. There are no non-trivial maps corresponding to cubic oscillators,
but we find a four-para\-meter family of such maps corresponding to
quartic oscillators. This seems to be the only such example.

}
\vfill\eject
\baselineskip=14pt
Suris [1] has found all one-degree-of-freedom Hamiltonian systems that
possess {\it integrable} discretizations of standard type.
If $f(x,0)$ is rational, it must be
a polynomial of degree $\le3$ and $f(x,\eps)$ is rational. Naive counting
suggests that if
some 2--d.o.f. continuous systems possess a second integral, then some
2--d.o.f. symplectic maps might possess one integral. Here we search for
such maps, which turn out to be rare: we find one four-parameter family
of maps with one integral, a three-parameter subset of which is rotationally
invariant and hence has a second integral; this subset corresponds
to discretizations of the two-parameter family of Hamiltonians
${1\over2}(p_1^2+p_2^2)+A(x_1^2+x_2^2)+B(x_1^2+x_2^2)^2$.
The rarity is perhaps because we want a whole family of integrable maps
(depending on $\eps$), not just an isolated map.

A symplectic map of standard type is written
$$ \x_{n+1}-2\x_n+\x_{n-1}=\eps\f(\x_n,\eps),\qquad
\f(\x,\eps)=\nabla V(\x,\eps)\eqno(1)$$
and the small parameter $\eps$ may be thought of as the square of the time-step
in a discretization of $\ddot\x=\f(\x,0)$.
Assume that $\f$ and $V$ have expansions
$$ \f(\x,\eps)=\sum_{j=0}^\infty\eps^j \f_j(\x),\qquad
V(\x,\eps)=\sum_{j=0}^\infty\eps^j V_j(\x)$$
We want an integral $\Phi(\x,\y,\eps)$ such that
$$\Phi(\x_{n+1},\x_n,\eps)=\Phi(\x_n,\x_{n-1},\eps)\qquad\hbox{ for all }n
\eqno(2)$$
and assume that
$$ \Phi(\y,\x,\eps)=\Phi(\x,\y,\eps)=\Phi_0(\x,\y)+\eps\Phi_1(\x,\y), $$
the first restriction following from the reversibility (under $n\mapsto-n$)
of (1).

Our construction follows that of Suris. Define $\u_n=\x_n-\x_{n-1}+\eps
\f(\x_n,\eps)/2$; then from (1) and (2) the condition for $\Phi$ to
be an integral becomes
$$ \Phi(\x,\x+\eps\f(\x,\eps)/2-\u,\eps)=\Phi(\x,\x+\eps\f(\x,\eps)/2+\u,\eps)
\eqno(3)$$
which may be expanded as a Taylor series in $\eps$. At order $\eps^0$, we get
$\Phi_0(\x,\y)=\varphi(\x-\y)$ (where $\varphi$ is necessarily even),
and at order $\eps$,
$$
(\nabla\varphi(\u))\cdot\f_0(\x)=\Phi_1(\x,\x-\u)-\Phi_1(\x,\x+\u).\eqno(4)$$
In the one-degree-of-freedom case,
Suris's method for solving the functional equation (3) is to:

\item{(i)} Differentiate the order $\eps$ term twice, the order $\eps^2$ term
once, and combine with the order $\eps^3$ term; this gives
six equations in the six unknowns $\Phi_1^{(i)}(x,x\pm u)$ ($i=1,2,3$)
which turn out to
have rank 5. The consistency
condition for these equations is separable in $x$ and $u$, and he obtains
$\varphi'''(u)/\varphi'(u)=c$, giving three {\it candidates} for $\varphi$:
$u^2/2$, $(1-\cos\omega u)/\omega^2$, and $(\cosh\omega u-1)/\omega^2$.
\item{(ii)} For each candidate, deduce the functional form of $\Phi_1$ from
(4).
\item{(iii)} Take the most general function of this form, substitute into
(3), and solve for $f(x,\eps)$.

In each case, the candidate for $\varphi$ did in fact give a family of
solutions for $f$. With two degrees of freedom, the same steps could
be followed: for (i), to get an overdetermined system it is necessary to
differentiate the order $\eps$ term three times, etc., proceeding to the
order $\eps^4$ term; this gives a rank-18 set of twenty equations in
twenty unknowns and two consistency conditions relating $\varphi^{(i,j)}(\u)$
and $\f^{(i,j)}(\x)$; these equations are complicated and probably
not separable. However, the severest restriction is the new one that
$\f(\x,\eps)$ must be a gradient:

\item{(iv)} Given $\Phi_1$, require ${\partial f_1\over\partial
x_2}\equiv{\partial f_2\over\partial x_1}$.

This must be checked for any proposed $\varphi$. We have taken
$\varphi(\u)=(u_1^2+u_2^2)/2$ and searched for solutions of (3) satisfying
(iv). From (4), $\Phi_1(\x,\y)$ must be a polynomial, and at
most quadratic in each variable, corresponding to a quartic potential $V_0$.
Hence we take
$$\Phi_1(\x,\y)=\sum_{i,j,k,l=0}^2 p_{ijkl} x_1^i x_2^j y_1^k y_2^l\qquad(
p_{ijkl}=p_{klij}\quad \forall i,j,k,l).$$
The consistency conditions on $\f$ when solved directly from (3) are
very complicated, so we determine them term-by-term. Once a solution at the
first few orders is obtained we go back and check (3) directly.
First determine $V_0$ from (4):
$$ \eqalign{V_0=-\big( &
  2\,x_2\,p_{0100} + {x_2^2}\,
    \left( p_{0101}+ 2\,p_{0200} \right)  +
   2\,{x_2^3}\,p_{0201} + {x_2^4}\,p_{0202} + \cr
   &x_1\,\left( 2\,p_{1000} +
      x_2\,\left( 2\,p_{1001} + 2\,p_{1100} \right)  +
      {x_2^2}\,\left( 2\,p_{1002} + 2\,p_{1101} \right)  +
      2\,{x_2^3}\,p_{1102} \right)  + \cr
   &x_1^2\,\left( p_{1010} + 2\,p_{2000} +
      x_2\,\left( 2\,p_{1110} + 2\,p_{2001} \right)  +
      {x_2^2}\,\left( p_{1111} + 2\,p_{2002} \right)  \right)  + \cr
   &x_1^3\,\left( 2\,p_{2010} + 2\,x_2\,p_{2011} \right)  +
   x_1^4\,p_{2020}\big)\cr
}$$
and then at order $\eps^n$ ($n=2,3,4$) we have
$$ \u.(\f_{n-1}+\g_n(\x).\f_{n-2}+\h_n(\x))={\bf 0}\eqno(5)$$
where $\g_n(\x)$ is a polynomial and $\g_2=0$. Solving (5) for $\f_{n-1}$
gives a consistency condition at that order and a value which can be
substituted into the next order. The consistency conditions take the
form of polynomials in $x_1$, $x_2$ (whose coefficients are functions of
the $p_{ijkl}$) which must be zero; this gives polynomial equations,
of degree $n$, in the $p$'s. To simplify the equations we took various
choices for $V_0$ and attempted to solve these consistency equations.
(The actual equations and solutions are derived using a symbolic manipulator.)
Without any other restrictions there are 14 equations at order 2, 27 at
order 3, and 38 at order 4.

If the map is to be completely integrable, then the vector field it
approximates as $\eps\to0$ must be also; hence we first tried $V_0$'s
corresponding to known integrable oscillators [2,3]:
$A x_1^2+B x_2^2-x_1^2x^2-2x_2^2$;
$A(x_1^2+16 x_2^2)-x_1^2x_2-16x_2^3/3$ (two cases of the H\'enon-Heiles
Hamiltonian); $x_1^2x_2+2x_2^3$; $x_1^4+12x_1^2x_2^2+16x_2^2$;
$x_1^4+6x_1^2x_2^2+8x_2^2$; and $A(x_1^2+x_2^2)+(x_1^2+x_2^2)^2$. Only
in the last case is there a consistent solution for the $p$'s (given
below), although
sometimes it is necessary to continue to order $\eps^4$ to reach the
inconsistency.

Now consider whether there are any maps with only {\it one} integral.
If so, then (4) shows (take $\partial/\partial\u$ and set $\u=0$)
that the integral must correspond to the Hamiltonian
as $\eps\to0$, i.e. $\Phi_1(\x,\x)=-V_0(\x)$.
We are not interested in the trivial solutions---those which give linear
maps, those which give uncoupled maps, or those which uncouple after
a linear symplectic change of variables. For the form
$$ V_0=A x_1^2 + B x_2^2 + \alpha x_1^3 + \beta x_1^2 x^2 + \gamma x_1 x_2^2
+\delta x_2^3$$
the only such solutions found had $A=B$ and $3(\alpha\gamma+\beta\delta)=
\beta^2+\gamma^2$; a long calculation shows that these maps all uncouple
after a linear symplectic change of variables. Thus there do not seem
to be any nontrivial maps of this form with one integral
corresponding to cubic oscillators.

For the form
$$ V_0=A x_1^2 + B x_2^2 + \alpha x_1^3 x_2 + \beta x_1 x_2^3 $$
we found no nontrivial solutions.

For the form
$$ V_0 = A x_1^2 + B x_2^2 + \alpha x_1^4 + \beta x_1^2x_2^2 + \gamma x_2^4$$
we found a four-parameter family of nontrivial solutions with
$\beta=2\alpha=2\gamma$ and arbitrary $A,B$. This seems to be the only
nontrivial solution: Let
$$ Q=b\left((1-c\eps)x_1^2+(1-d\eps)x_2^2\right).$$
Then
$$\eqalign{
V=-{2+a\eps\over b\eps^2}\ln\big(2(1-c\eps)&(1-d\eps)-\eps Q\big)-
{x_1^2+x_2^2\over\eps} \cr
f_1=x_1 {(a+2d)(1-c\eps)+Q\hskip 1.5em \over (1-d\eps)(1-c\eps)-\eps
Q/2},&\qquad
f_2=x_2 {(a+2c)(1-d\eps)+Q\hskip 1.5em \over (1-d\eps)(1-c\eps)-\eps Q/2}\cr
} \eqno(6)$$
with integral
$$
-2\Phi_1=a(x_1y_1+x_2y_2)+b(x_1y_1+x_2y_2)^2/2+c(x_2^2+y_2^2)+d(x_1^2+y_1^2)$$

When $c=d$, the potential $V$ is rotationally invariant and we expect a second
integral; the method determines this automatically, for in this case one
finds a term $e(x_1y_2-x_2y_1)^2$ in $\Phi_1$ with $V$ independent of $e$.
The obvious extension of this map to $n$ degrees of freedom is also
integrable:
$$ \eqalign{\f&={a+2c+b|\x|^2\over1-c\eps-\eps b|\x|^2/2}\,\x \cr
 V&=-{2+a\eps\over b\eps^2}\ln\left(2-b\eps|\x|^2\right)-{|\x|^2\over\eps}\cr
V_0&= {a+2c\over2}|\x|^2 + {b\over2}|\x|^4 \cr
} \eqno(7)$$

Grammaticos et al. [4] have proposed an integrability test for maps: movable
singularities should not propagate in time, and memory of the initial condition
should survive the singularity. To apply the test (for simplicity, for
2 degrees of freedom), write the map in the
form $\x_{n+1}=-x_{n-1}+\alpha/(1-\beta^2|\x_n|^2)\x_n$ and take initial
conditions $\x_0$ arbitrary, $\x_1=(x_{11},0)$ (this suffices because we
can rotate coordinates). Compute the next three iterates and then let
$x_{11}\to1/\beta$; this gives the iterates
$$ (x_{00},x_{01}),\quad (1/\beta,0),\quad (\infty,-x_{01}),\quad
(-1/\beta,0),\quad (-x_{00},x_{01})$$
showing that the proposed integrability test is satisfied here.

We can now use the new ``angular momentum'' integrals to reduce the map to
one degree of freedom. This illustrates that the two processes,
(i) forming an integrable map approximating a continuous system and
(ii) reducing by the rotational symmetry, do not commute.
Rotate coordinates so that
$x_i=y_i=0$ for $i>2$. Write the map as
$$\eqalign{\x'=&\,\x+\p \cr
	   \p'=&\,\p+\x' g(|\x'|)}$$
and define new variables (chosen to correspond to the continuous case)
$$ \eqalign{L=&\,x_1p_2-x_2p_1\cr
	    (r,\theta):\quad& \hbox{polar coordinates for }(x_1,x_2)\cr
	    p_r = &\,(x_1 p_1 + x_2 p_2)/r\cr
	    p_\theta=&\,L\cr}
$$
in which the map can be written (after much algebra)
$$\eqalign{
	(r^2)'&={L^2\over r^2}+(p_r+r)^2\cr
	(rp_r)'&={L^2\over r^2}+rp_r+p_r^2+(r')^2g(r')\cr
	       &=(r')^2(1+g(r'))-r(r+p_r)\cr
	L'&=L\cr
	\theta'&=\theta+\sin^{-1}{L\over r r'}\cr
  }\eqno(8) $$
with integral
$$ {L^2\over r^2}+p_r^2-\eps\left(r(r+p_r)\left(a+br(r+p_r)/2\right) +c\left(
{L^2\over r^2}+(r+p_r)^2+r^2\right)\right)
$$
However, starting with the continuous system corresponding
to (7) ($\ddot\x=\nabla V_0(|\x|)$) and performing the corresponding reduction
gives
$$ \ddot r={L^2\over r^3}+V_0'(r),\qquad\dot\theta={L\over r^2}$$
which does {\it not} have an integral discretization of standard type. So
we have extended Suris's list of systems possessing integrable discretizations
by one;
the catch is the our reduced system (8) is not of standard type. This suggests
that there may be rational integrable discretizations of any rational
Hamiltonian if one enlarges the allowed class of maps.

My thanks to James Meiss for pointing out [1] and for many helpful discussions.
\bigskip\bigskip
\noindent{\bf References}
\bigskip
\baselineskip=12pt
\parskip=5pt
\leftskip 2.5em \parindent = -2.5em
[1] Suris, Yu. B. (1989) Integrable mappings of the standard type, {\sl Funct.
Anal. Appl. \bf 23}(1), 74--75.

[2] Ramani, A., Grammaticos, B., and Bountis, T. (1989) The Painlev\'e
property and singularity analysis of integrable and nonintegrable
systems, {\sl Physics Reports \bf 180}(3), 159--245; (see in
particular pp. 185--188).

[3] Bountis, T., Segur, H., and Vivaldi, F. (1982) Integrable Hamiltonian
systems and the Painlev\'e property, {\sl Phys. Rev. A \bf 25}(3), 1257--1264.

[4] Grammaticos, B., Ramani, A., and Papageorgiou, V. (1991) Do
integrable mappings have the Painlev\'e property?. {\sl Phys. Rev. Lett. \bf
67}
(14), 1825--1828.
\vfill\eject\end